\documentclass[runningheads]{llncs}
\usepackage[T1]{fontenc}
\usepackage{graphicx}
\usepackage{subcaption}
\usepackage{multirow}
\usepackage{tabularx}
\usepackage{makecell}
\usepackage{booktabs}
\usepackage{url}

\begin{document}
\title{Usability Analysis of Configurator User Interfaces with Multimodal Large Language Models}
\titlerunning{Usability Analysis of Configurator UIs with Multimodal LLMs}
%

\author{Sebastian Lubos\inst{1} \and
Alexander Felfernig\inst{1} \and
Damian Garber\inst{1} \and
Adnan Kraljić\inst{1} \and
Tarik Kraljić\inst{1} \and
Viet-Man Le\inst{1} \and
Thi Ngoc Trang Tran\inst{1} \and
Gerhard Leitner\inst{2} \and
Julian Schwazer\inst{2} \and
Doris Suppan\inst{3}  \and
Reinhard Willfort\inst{3} \and
Ivan Dukic\inst{4} \and
Jeremias Fuchs\inst{4} \and
Manuel Henrich\inst{5}
}

\authorrunning{S. Lubos et al.}

\institute{Graz University of Technology, Inffeldgasse 16b/II, 8010 Graz, Austria
\email{sebastian.lubos@tugraz.at} \and
University of Klagenfurt, Universitätsstraße 65-67, 9020 Klagenfurt, Austria \and
isn – innovation service network GmbH,  Grabenstraße 231, 8045 Graz, Austria \and
Morgendigital GmbH, Jahnstraße 18, 6020 Innsbruck, Austria \and
UNiQUARE Software Development GmbH, Lannerweg 9, 9201 Krumpendorf, Austria
}

%
%
\maketitle              
\begin{abstract}
Configuration is a key technology for tailoring complex software systems, services, and products. A successful application of configurators not only depends on technical correctness, performance, and domain modeling but also on their \textit{usability}. While general usability heuristics are widely used, configurator-specific criteria and tool support for systematic \textit{user interface (UI)} analysis are limited. This paper explores the use of \textit{multimodal large language models (MLLMs)} for scalable and semi-automated usability analysis of configurator UIs. We synthesize 18 configurator-specific usability criteria from the literature and apply these criteria in an MLLM-based analysis of 16 real-world configurators. Each criterion is assessed individually to generate severity ratings for usability issues and actionable improvement suggestions. A review of the results confirms that MLLMs can reliably identify configurator-specific usability issues and provide domain-aware improvement recommendations. Although human validation remains necessary, this approach has the potential to significantly reduce the required effort to analyze configurator usability.

\end{abstract}

\section{Introduction} \label{sec:introduction}
Configurators are key components in software product lines and mass customization~\cite{felfernig2007standardized,felfernig2014knowledge}. They enable customers to tailor complex software, services, and products to individual user needs and often serve as the primary entry point for placing orders~\cite{abbasi2013anatomy}. However, even if a configurator produces technically correct configurations, it may still fail to meet the expectations of its intended users~\cite{rabiser2012qualitative}. Since customization is often performed by end users with limited domain expertise, the \textit{usability} of configurator \textit{user interfaces (UIs)} has to be assured~\cite{leitner2014userinterfaces}.

Configuration-related research often focuses on the aspects of correctness, performance, and domain knowledge modeling~\cite{felfernig2024analysis,siegmund2012predicting,thum2014classification}. However, the UI, as the primary interaction layer with end users, receives less attention. Configurator UIs help users to understand available options, express preferences, explore alternatives, construct valid solutions, and complete orders~\cite{leitner2014userinterfaces}. These functionalities are highly relevant for the successful completion of configuration sessions.

According to \textit{ISO~9241-11:2018}, \textit{usability} is the extent to which users achieve specified goals effectively, efficiently, and satisfactorily~\cite{iso2018iso9241}. In \textit{human–computer interaction (HCI)}, usability is a central design and evaluation concern~\cite{hewett1992hci}. To analyze the usability of software systems, usability heuristics, such as \textit{Nielsen’s heuristics}~\cite{nielsen1994enhancing}, help to reveal potential usability issues. However, these heuristics address general issues but not configurator-specific challenges, such as large option spaces or complex compatibility constraints. Also, heuristic analyses require substantial manual effort and expertise. The lack of configurator-specific criteria and related tool support~\cite{abbasi2013anatomy} is a probable reason for the absence of a systematic evaluation of configurator usability in practice.

Recent advances in \textit{multimodal large language models (MLLMs)}~\cite{yin2024survey} offer new opportunities for scalable and semi-automated usability analysis. MLLMs can process textual descriptions alongside visual representations. Initial studies report promising results in general usability evaluations~\cite {guerino2025can,lubos2026investigating,pourasad2025does,zhong2025synthetic}, but their potential for assessing configurators remains unexplored.

This paper addresses this gap. We review existing work to derive configurator-specific usability criteria and apply them in an MLLM-based analysis of real-world configurators. We aim to understand whether a state-of-the-art MLLM can accurately detect configurator-specific issues and suggest related improvements.

The contributions of this paper are the following:
\begin{itemize}
    \item We present a set of configurator-specific usability criteria grounded in existing research.
    \item We introduce an MLLM-based usability-analysis approach applied to real-world configurators.
    \item We evaluate the strengths, limitations, and implications of our approach for improving configurator UIs.
\end{itemize}

\section{Usability Criteria for Configurators} \label{sec:usability}
Multiple guidelines describe rather general criteria for good usability in interactive software systems, including the \textit{ISO 9241-11:2018} standard \cite{iso2018iso9241}, \textit{Nielsen heuristics}~\cite{nielsen1994enhancing}, and platform-dependent guidelines such as the \textit{Human Interface Guidelines} by \textit{Apple}\footnote{\url{https://developer.apple.com/design/human-interface-guidelines}} or the \textit{UI Design} best practices for \textit{Android}\footnote{\url{https://developer.android.com/design/ui}}. While these criteria are also applicable to configurator UIs, they lack configurator-specific aspects. As no guideline considers the specific requirements of configurator users, we summarize the related literature on this topic and extract relevant usability criteria for configurators.

To identify relevant studies, we conducted keyword searches in \textit{Google Scholar}\footnote{\url{https://scholar.google.com}} and the proceedings of software engineering conferences in the \textit{ACM Digital Library}\footnote{\url{https://dl.acm.org}}, using related terms such as ``\textit{configurator usability}'', ``\textit{configurator user interfaces}'', and ``\textit{interactive product configurator}''. We screened the retrieved papers for relevance and applied backward snowballing by reviewing their references to find additional relevant studies. While this literature review was not systematic, it provides a solid basis for deriving configurator-specific usability criteria grounded in existing research.

Rogoll and Piller \cite{rogoll2004product} compared 12 configurators for mass customization in the apparel industry from the user (customer) perspective to identify good and bad practices. The authors emphasize that configurators must fulfill three major tasks for customers. Firstly, \textit{risk reduction and trust building}, which is related to different aspects of uncertainty on the customer side. This involves a lack of knowledge about which solution fulfills their needs and what their needs actually are. Additionally, users do not have the opportunity to evaluate the configuration before placing their order. Secondly, the \textit{visualization} of the configuration can directly mitigate this uncertainty by illustrating the design of the software, service, or product through drafts, renderings, or descriptions. Thirdly, the \textit{usability} is directly responsible for the success or failure of a configurator, as it determines whether the configuration process can be completed successfully or will be aborted. Overall, a configurator must support users and reduce their uncertainty by building confidence and demonstrating competence, which in turn increases customer trust and the likelihood of ordering.

Rabiser et al. \cite{rabiser2012qualitative} reviewed the features of academic and commercial configurators to identify key capabilities that guide users based on their needs. These identified features were implemented in a configurator, which was evaluated in a user study to derive general implications for tool developers. Related insights involve measures to tackle complexity challenges, for example, by hiding certain options depending on the current configuration step or by enabling option filtering. Additionally, navigation is supported by guiding users through the configuration in a predefined yet flexible process, allowing them to choose arbitrary configuration options at any time to avoid a feeling of forced selection. Also, the state of configuration should be immediately reflected by visualizing changes and ensuring the validity of the configuration throughout the process. Explanations help users to understand the process, while suggestions for resolving problems help them to comprehend and address dependency errors.

Abbasi et al. \cite{abbasi2013anatomy} reviewed 111 real-world web sales configurators from various industries to highlight good and bad practices. Based on their review, they suggested guidelines for the efficient engineering of configurators and considered three essential dimensions, including \textit{rendering configuration options}, \textit{constraint handling}, and \textit{configuration process support}. They highlight the need for self-explanatory configurators that provide clear guidance during the configuration, including explanations of constraints and transparent communication of changes when dependencies are propagated. Additionally, they argue for the need of immediate consistency checks and flexible stateful navigation, which allows users to control the configuration process and keep selected properties unchanged.

Trentin et al. \cite{trentin2013sales} discuss the capabilities of sales configurators that help to avoid the \textit{product variety paradox}, i.e., the risk that offering more product variety and customization results in decreased sales. These capabilities include: \textit{focused navigation}, \textit{flexible navigation}, \textit{benefit-cost communication}, \textit{easy comparison}, and \textit{user-friendly product space description}. \textit{Focused navigation} is about supporting an efficient configuration process that narrows down the search space fast. If users can start with options they are most certain about, they can invest more time in learning and deciding about uncertain properties. \textit{Flexible navigation} aims to minimize the effort required to modify a configuration by allowing changes to previous steps without losing progress. \textit{Benefit-cost communication} should explain the consequences of the available choice options in terms of what the customer gets and in terms of what the customer has to give (monetary and nonmonetary costs, e.g., delivery time). This helps to assess the utility of a configuration. Providing \textit{easy comparison} of configured variants supports more efficient decision-making. A \textit{user-friendly product space description} should focus on the needs and abilities of different potential customers and contexts of use. The positive effect of these capabilities on the creative and hedonic value perceived by users of the configurator was confirmed in a user study \cite{perin2013effect}.

Leitner et al. \cite{leitner2014userinterfaces} summarize five key design principles for configurator UIs: (\textit{i}) \textit{Customize the customization process} by personalizing the configuration environment based on the background of the current user, for example, by providing a \textit{parameter-based interface} for direct configuration of technical properties by domain experts, and a \textit{needs-based interface} for non-experts that translates customer requirements into technical properties. (\textit{ii}) \textit{Provide starting points} to enable users to begin with predefined configurations, and not only from scratch. (\textit{iii}) \textit{Support incremental refinement} by highlighting tradeoffs between properties and alternatives to support users in identifying their preferences during the configuration process. (\textit{iv}) \textit{Exploit prototypes to avoid surprises} by visualizing the configuration. (\textit{v}) \textit{Teach the consumer} to expand their domain knowledge by explaining parameters, highlighting the constraints, and explaining conflicting requirements if no solution can be found.

Leclercq et al. \cite{leclercq2018studying} reviewed car configurators, focusing on HCI principles and design flaws to identify the most frequent usability problems in these systems. Besides the violation of basic HCI principles (e.g., information overload, missing feedback), the most frequent problem was the inappropriate implementation of configurator-specific functionalities. Examples of identified issues were related to the \textit{navigation} within the configurator and included missing state information, unrecognizable controls to proceed, and a lack of visualization. The configuration itself lacked validity checks, which allowed incompatible user selections. Transparent explanations and suggestions to repair errors were missing. 

A user study by Leclercq et al. \cite{leclercq2022essential} identified that product validity and adequacy have the highest priorities for users of configurators. Usability and real-time visualization were the most important factors. Based on these insights, the authors suggest corresponding UI practices to support these priorities. These include systematically \textit{organizing large configuration spaces} to avoid crowded UIs if many options are available, \textit{avoiding disabled configuration options}, \textit{notifying users} by highlighting changes and communicating dependencies, \textit{assisting and teaching} users appropriately by explaining the impact of choices, \textit{providing parametrized product visualization} by offering multiple visualization options, and \textit{customizing the configuration process} to support different user expectations.

Building on these foundations, we summarize the usability criteria that are relevant to achieving a good user experience with configurators. We organize these criteria into four categories, summarized in Tables~\ref{tab:criteria_configuration}--\ref{tab:criteria_visualization}. For all criteria, we provide a description that can be used to evaluate the extent to which a configurator application fulfills the criterion.

The criteria in Table~\ref{tab:criteria_configuration} summarize aspects related to the \textit{configuration process} itself, in which users define their preferences and specify the relevant options. This involves aspects related to the presentation and selection of configuration options, comparison of alternatives, and prevention of invalid configurations. 

\begin{table}[h!]
    \vspace{-8pt}
    \centering
    \caption{Evaluation criteria of the \textit{configuration process} category.}
    \label{tab:criteria_configuration}
    \vspace{4pt}
    \begin{tabular}{@{}cp{2.4cm}p{7.75cm}p{1.3cm}@{}}
    \toprule
    \textbf{ID} & \textbf{Criterion} & \textbf{Description} & \textbf{Refs.} \\
    \midrule
    C1 & \makecell[tl]{Customized\\options} & Does the configurator adapt the available options to meet different user profiles (e.g., expert vs. non-expert), such as presenting a needs-based view or a parameter view, with a clear way to select the profile? & \cite{leclercq2022essential,leitner2014userinterfaces,trentin2013sales} \\
    C2 & \makecell[tl]{Organized\\configuration\\space} & Does the configurator help users with large option spaces by utilizing explicit mechanisms (e.g., grouping, search/filtering, multi-step configuration) to keep the number of simultaneously visible choices manageable? & \cite{leclercq2022essential,rabiser2012qualitative} \\
    C3 & \makecell[tl]{Availability\\of options} & Can users revisit and change previously set options without losing their current configuration state? & \cite{abbasi2013anatomy,leclercq2022essential,rabiser2012qualitative} \\
    C4 & \makecell[tl]{Auto-completion} & Does the configurator offer user-triggered auto-completion that fills in the remaining required options with defaults? & \cite{abbasi2013anatomy} \\
    C5 & \makecell[tl]{Variant\\comparison} & Can users keep multiple variants and compare them (e.g., ranking by properties, side-by-side, or highlighting differences) to review trade-offs? & \cite{leitner2014userinterfaces,trentin2013sales} \\
    C6 & \makecell[tl]{Error prevention} & Does the configurator prevent users from ending up with an invalid configuration (e.g., by disabling incompatible options, auto-resolving conflicts, or requiring conflict resolution before proceeding)? & \cite{abbasi2013anatomy,leclercq2018studying,rabiser2012qualitative} \\
    \bottomrule
    \end{tabular}
    \vspace{-8pt}
\end{table}

Table~\ref{tab:criteria_explanation} summarizes criteria related to \textit{explanations} within the system, which help users to understand the product domain, dependencies, and give support if errors occur.  
In Table~\ref{tab:criteria_navigation}, criteria related to the \textit{navigation} within the configurator are summarized. These criteria assess whether users can navigate efficiently, transparently, and flexibly, based on their individual needs. 
Table~\ref{tab:criteria_visualization} collects criteria regarding the \textit{visualization} of the configuration, which help users to evaluate whether a configuration adequately fulfills their needs.

\begin{table}[h!]
    \centering
    \caption{Evaluation criteria of the \textit{explanation} category.}
    \label{tab:criteria_explanation}
    \vspace{4pt}
    \begin{tabular}{@{}cp{2.4cm}p{7.75cm}p{1.3cm}@{}}
    \toprule
    \textbf{ID} & \textbf{Criterion} & \textbf{Description} & \textbf{Refs.} \\
    \midrule
    E1 & \makecell[tl]{Providing\\domain\\knowledge} & Does the configurator provide in-context explanations of options (e.g., tooltips or examples) that clarify meaning and help users make informed choices? & \cite{abbasi2013anatomy,leitner2014userinterfaces,rogoll2004product,trentin2013sales} \\
    E2 & \makecell[tl]{Transparency\\of dependencies} & Does the configurator explain dependencies of options at decision time (e.g., ‘choosing X requires Y’, ‘this removes Z’, impact on price/delivery) to highlight consequences of choices? &\cite{leclercq2018studying,leitner2014userinterfaces,trentin2013sales} \\
    E3 & \makecell[tl]{Transparency\\of errors} & Does the configurator explain why a configuration is inconsistent (e.g., highlighting which selected options are in conflict) using actionable, non-technical language? & \cite{abbasi2013anatomy,leclercq2022essential,leclercq2018studying,leitner2014userinterfaces,rabiser2012qualitative} \\
    E4 & \makecell[tl]{Repair\\suggestions} & Does the configurator suggest actionable repair options (e.g., ‘change X to Y or Z’ or ‘remove X’) to assist users in resolving errors with inconsistent constraints? & \cite{abbasi2013anatomy,leitner2014userinterfaces,rabiser2012qualitative} \\
    \bottomrule
    \end{tabular}
\end{table}

\begin{table}[h!]
    \centering
    \caption{Evaluation criteria of the \textit{navigation} category.}
    \label{tab:criteria_navigation}
    \vspace{4pt}
    \begin{tabular}{@{}cp{2.4cm}p{7.75cm}p{1.3cm}@{}}
    \toprule
    \textbf{ID} & \textbf{Criterion} & \textbf{Description} & \textbf{Refs.} \\
    \midrule
    N1 & \makecell[tl]{Focused\\navigation} & Does the configurator provide a clear, task-aligned sequence of steps that helps users reach a valid result efficiently (e.g., using a relevant order of decisions or avoiding unnecessary steps)? & \cite{abbasi2013anatomy,leclercq2018studying,rabiser2012qualitative,rogoll2004product} \\
    N2 & \makecell[tl]{Manual step\\transition} & Does the configurator support moving forward/backward using consistently placed, clearly labeled controls? & \cite{abbasi2013anatomy,rabiser2012qualitative} \\
    N3 & \makecell[tl]{Flexible\\navigation} & Can users modify or undo earlier selections without losing the current configuration state?  & \cite{abbasi2013anatomy,rabiser2012qualitative,trentin2013sales} \\
    N4 & \makecell[tl]{Progress\\indication} & Does the configurator clearly indicate the current step, remaining steps, and completion status (e.g., stepper or progress bar)? & \cite{abbasi2013anatomy,leclercq2018studying} \\
    N5 & \makecell[tl]{Variant\\persistence} & Can users save, name, or restore previous full configuration variants during or after the session (e.g., version history or saved configurations)? & \cite{trentin2013sales} \\
    N6 & \makecell[tl]{Starting points} & Does the system offer practical starting points to support different user needs (e.g., starting from predefined configurations or selecting properties to start with)? & \cite{leitner2014userinterfaces,trentin2013sales} \\
    \bottomrule
    \end{tabular}
\end{table}

\begin{table}[h!]
    \centering
    \caption{Evaluation criteria of the \textit{visualization} category.}
    \label{tab:criteria_visualization}
    \vspace{4pt}
    \begin{tabular}{@{}cp{2.4cm}p{7.75cm}p{1.3cm}@{}}
    \toprule
    \textbf{ID} & \textbf{Criterion} & \textbf{Description} & \textbf{Refs.} \\
    \midrule
    V1 & \makecell[tl]{Product\\preview} & Does the configurator provide a product preview that is continuously updated after changes to reflect the current configuration? & \cite{abbasi2013anatomy,leclercq2018studying,leitner2014userinterfaces,rabiser2012qualitative,rogoll2004product} \\
    V2 & \makecell[tl]{Customized\\preview} & Can users switch between preview modes (e.g., 2D/3D images or a textual summary), with a consistent relation to the current configuration? & \cite{leclercq2022essential,rogoll2004product} \\
    \bottomrule
    \end{tabular}
\end{table}

Overall, these usability criteria encompass a broad range of aspects that can help to improve the quality of user interaction with configurators. While these criteria can be used for the manual evaluation of configurator UIs, we utilize them to conduct an automated usability analysis with MLLMs to identify usability issues and generate explicit improvement suggestions.

Some of the usability criteria are tightly coupled to specific features of a configurator, e.g., the \textit{auto-completion} of configurations (see C4 in Table~\ref{tab:criteria_configuration}). However, they can be evaluated beyond identifying whether these features are present by considering how intuitive they are to use. In this sense, an issue can still be present if a feature is available but does not sufficiently support users in their interactions. This aligns with \textit{ISO~9241-11:2018}~\cite{iso2018iso9241}, which defines \textit{usability} as the extent to which users achieve specified goals effectively, efficiently, and satisfactorily. While general usability criteria, such as \textit{Nielsen heuristics}~\cite{nielsen1994enhancing}, remain helpful for identifying usability issues in configurators, combining them with domain-specific criteria helps to address the specific user needs for specialized configurator applications.

\section{MLLM-based Usability Analysis} \label{sec:improvement}
Initial approaches for automated usability analysis used AI-based methods, such as heuristic checkers and rule-based approaches~\cite{castro2022automated,namoun2021review}. These are typically limited in terms of the considered evaluation aspects and context-dependent issues~\cite{kuric2025systematic}. More recently, \textit{multimodal large language models (MLLMs)}, i.e., models that combine textual and visual understanding ~\cite{yin2024survey}, have shown new possibilities for usability analysis. 
The general idea is to provide screenshots of an application to be analyzed, together with instructions and specific evaluation criteria, to assess how well these criteria are satisfied~\cite{duan2024generating,guerino2025can,lubos2025towards,lubos2026investigating,pourasad2025does,zhong2025synthetic}. In many cases, the analysis generated by the MLLM had good overlap with assessments of usability experts. Nevertheless, their role remains supportive, and UI experts are still needed for further evaluation and validation~\cite{guerino2025can,lubos2026investigating,pourasad2025does,zhong2025synthetic}. 

Up to now, most studies have focused on general usability criteria, for example \textit{Nielsen's heuristics}~\cite{nielsen1994enhancing}, for the analysis of different user interfaces, including design mockups~\cite{duan2024generating}, mobile apps~\cite{pourasad2025does,zhong2025synthetic}, and web applications~\cite{guerino2025can,lubos2026investigating}. The usage of application-specific usability criteria was investigated in~\cite{lubos2025towards}, where the analysis was tailored to specific criteria for recommender systems. It showed promising results in identifying and explaining related usability issues in user interfaces of recommender systems.

In contrast to prior work, this work focuses specifically on \textit{configurator user interfaces} and on usability criteria that reflect the specific challenges users face when interacting with them. Furthermore, we use screen recordings as the evaluation context, whereas previous studies rely on static screenshots~\cite {guerino2025can,lubos2025towards,lubos2026investigating,pourasad2025does,zhong2025synthetic}. Using videos instead of screenshots helps preserve interaction dynamics and UI state transitions. This enables a more comprehensive usability analysis. In line with previous work, we see significant potential in applying MLLMs to analyze configurator user interfaces and identify relevant usability issues. An automated analysis can potentially reduce the manual effort for these evaluations.

The underlying idea is to provide an MLLM with all the data needed to analyze the usability of a configurator user interface. This includes example user interactions, specified evaluation criteria, and explicit task instructions. For the configurator application, a representative screen recording of a user interaction is provided as visual context. To include the evaluation criteria, we use the descriptions of the usability criteria described in Tables~\ref{tab:criteria_configuration}--\ref{tab:criteria_visualization}. While it would be possible to evaluate multiple criteria at once, we consider one criterion per prompt to ensure more focused and precise results. 

\begin{figure}
    \centering
    \includegraphics[width=0.99\linewidth]{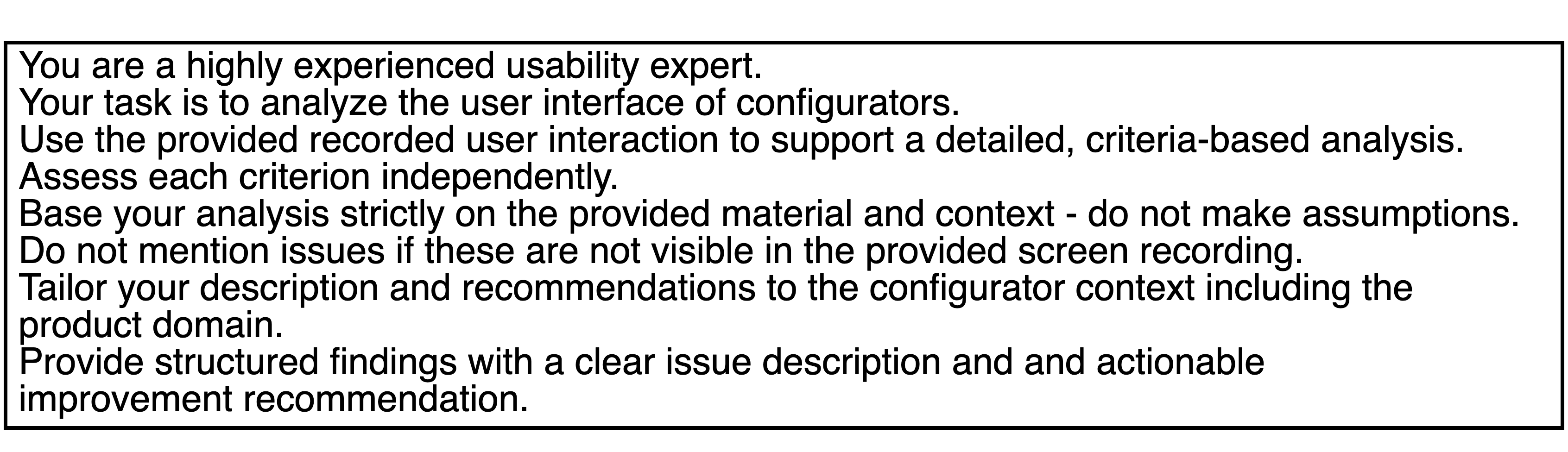}
    \caption{System prompt for the usability analysis, providing general instructions.}
    \label{fig:system_prompt}
\end{figure}

\begin{figure}
    \centering
    \includegraphics[width=0.99\linewidth]{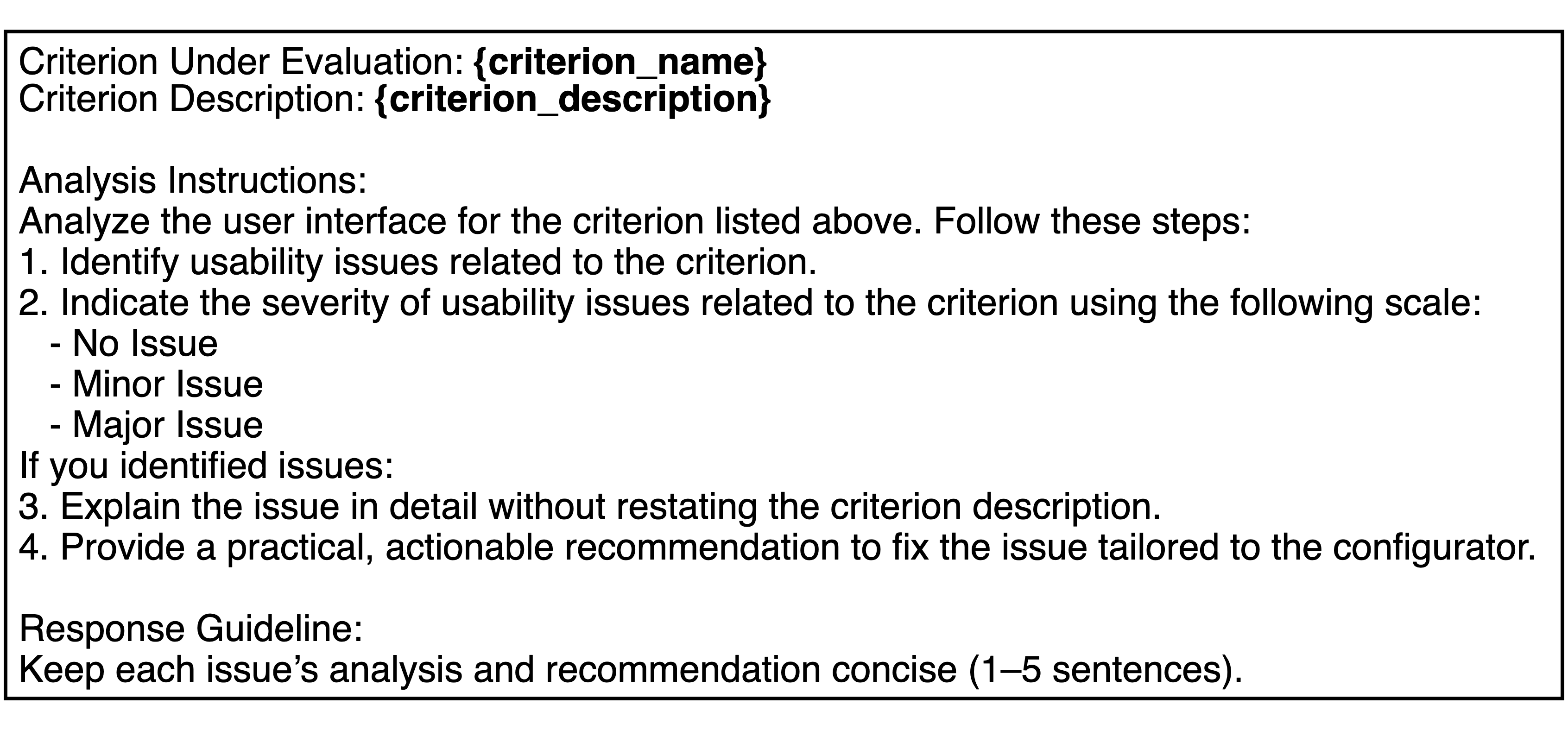}
    \caption{Prompt template for the usability analysis, where variable placeholders are indicated in braces ``\textbf{\{\}}''. Result formatting instructions are omitted.}
    \label{fig:prompt}
\end{figure}

The task instruction provides detailed guidance and clearly defines the expected output. In many cases, the usability analysis of criteria is not a binary decision, but rather identifies issues with different levels of severity. To account for this, the MLLM is instructed to provide a \textit{severity rating} on a 3-point scale (\textit{no issue}, \textit{minor issue}, \textit{major issue}). Additionally, if an issue is identified, a description of the issue and an explanation of how it can be improved are requested. We split the instructions for the MLLM into a \textit{system prompt}, which includes general instructions for the usability analysis (see Figure~\ref{fig:system_prompt}), and a \textit{user prompt} template that defines the context for the current criterion (see Figure~\ref{fig:prompt}).

\section{Methodology} \label{sec:methodology}

\subsection{Dataset} \label{ssec:dataset}
To evaluate our approach with real-world configurators, we selected \textit{16 examples} from the \textit{Cyledge configurator database}~\cite{cyledge2022configurator}. We selected one configurator from each available industry category to obtain a broad sample across domains. Each configurator had to be publicly accessible, available in English, and still online at the time of the study.

For each configurator, one of the authors interacted with the system\footnote{Using Google Chrome v143 on macOS v26} to demonstrate its functionality and interaction flows. Although these recordings do not reflect full, realistic customer journeys, they illustrate the types of actions and interface elements that real users commonly engage with when configuring and ordering a product. As a basis for our usability analysis, we recorded these interactions as full-screen MP4 videos. Compared to static application screenshots, a screen recording captures interaction sequences and transitions, which provides a richer context for analysis. The selected configurator applications are summarized in Table~\ref{tab:samples}, including the duration of each recorded interaction.

\begin{table}[h!]
    \vspace{-16pt}
    \centering
    \caption{Configurators used for the usability analysis, including the industry category, name, and duration of the recorded interaction in minutes.}
    \label{tab:samples}
    \vspace{4pt}
    \begin{tabular}{@{}cllcl@{}}
    \toprule
    \textbf{ID} & \textbf{Industry} & \textbf{Name} & \textbf{Duration} & \textbf{URL} \\
    \midrule
    1 & Accessories & Tie Creators & 05:22 & \url{tiecreators.com}\\
    2 & Apparel & Clothoo & 05:47 & \url{clothoo.com} \\
    3 & Beauty \& Health & eSalon & 04:19 & \url{esalon.com} \\
    4 & Electronics & AimControllers & 08:44 & \url{aimcontrollers.com} \\
    5 & Food \& Packaging & Oreo & 04:44 & \url{oreo.com} \\
    6 & Footwear & DIS & 05:49 & \url{designitalianshoes.com/en} \\
    7 & Games \& Music & Fender & 04:17 & \url{fender.com} \\
    8 & House \& Garden & Ergohide & 03:41 & \url{ergohide.com} \\
    9 & Industrial Goods & Altrex & 01:23 & \url{altrex.com/en} \\
    10 & Kids \& Babies & Nuk & 04:02 & \url{nuk.de} \\
    11 & Motor Vehicles & Aixam & 02:49 & \url{aixam.com} \\
    12 & Office \& Merchandise & Austrian Post & 03:20 & \url{onlineshop.post.at/en-AT} \\
    13 & Paper \& Books & Packhelp & 07:39 & \url{packhelp.com} \\
    14 & Pet Supplies & 4Pets & 01:13 & \url{4pets-products.com/en} \\
    15 & Printing Platforms & Namemaker & 01:32 & \url{namemaker.com} \\
    16 & Sportswear \& Equipment & Aurum Bikes & 02:15 & \url{aurumbikes.com} \\
    \bottomrule
    \end{tabular}
\end{table}

\subsection{Model Selection} \label{ssec:model}
For the usability analysis, we used Google’s \textit{gemini-2.5-flash} MLLM~\cite{geminiteam2024gemini}. This model was suitable for evaluating recorded user interactions, as it supports the efficient processing of multimodal inputs, including video. Although larger models are available, we decided to use this model, as prior work indicated that larger models do not necessarily yield more precise results for usability evaluation tasks~\cite{lubos2026investigating}. Other models may produce different outcomes, and a performance comparison could reveal such differences. However, this paper focuses on demonstrating the feasibility of MLLM-based usability analysis for configurators, rather than optimizing model choice. A systematic comparison is out of scope.

We accessed the model in \textit{Python} via the Gemini Developer API.\footnote{\url{https://ai.google.dev/gemini-api/docs}} To obtain more deterministic outputs, we set the \textit{temperature} to $0.0$. The screen recordings were passed directly to the model using the default settings, which extract one frame per second.\footnote{\url{https://ai.google.dev/gemini-api/docs/video-understanding}} This choice balances \textit{cost} and \textit{detail}, as fewer frames reduce token usage\footnote{Each extracted frame contributes 258 tokens under default settings.} and speed up analysis, but involve the risk of omitting relevant interactions, whereas more frames increase detail at higher cost.

Context-token limits could become a problem for excessively long recordings, though the practical risk is low. With the default sampling rate, \textit{gemini-2.5-flash} can process videos of at least one hour before reaching the token limit, which is sufficient for typical user interactions. However, if the limit is exceeded, a lower-quality tokenization or a lower frame sampling rate could be applied, or recordings could be split into smaller segments.

The full implementation of the usability analysis, dataset with video samples, and complete analysis and evaluation results are available in our repository.\footnote{\url{https://anonymous.4open.science/r/configurator-usability-analysis-2206/}}

\subsection{Study design} \label{ssec:study_design}
We used the recorded samples from the dataset (see Section~\ref{ssec:dataset}) as input for the MLLM and obtained analysis results for the predefined usability criteria following the procedure outlined in Section~\ref{ssec:model}. In the following, we present the MLLM-based analysis results descriptively and discuss detailed evaluation results. This study concentrates on the \textit{plausibility} of MLLM-identified usability issues and related improvement recommendations. This aims to assess the practical utility of this approach. 

To assess the correctness and plausibility of the identified issues and suggested improvements, we engaged six reviewers with several years of academic and professional experience in software engineering to evaluate the analysis results. The descriptions of each MLLM-identified usability issue and suggested improvement were independently evaluated by exactly three reviewers. For every sample, the reviewers watched the screen recording and assessed the generated outputs against the relevant usability criterion using two binary judgments: 
\begin{itemize}
    \item \textit{Issue description plausibility}: Does the described issue constitute a genuine usability problem related to the criterion and the given configurator?
    \item \textit{Improvement recommendation plausibility}: Is the proposed improvement appropriate to fix the identified issue?
\end{itemize}
We aggregate the three independent judgments by \textit{majority vote} and report the \textit{inter-rater reliability} to quantify agreement among reviewers.

\section{Results} \label{sec:results}
In total, 288 MLLM-based usability analyses were generated by evaluating 18 usability criteria across 16 configurators. In 140 cases, the results indicated \textit{no issue}, while 73 \textit{minor issues} and 75 \textit{major issues} were identified.

The execution times of the MLLM-based analysis ranged from 3.4 to 52.8 seconds per criterion (14.4 seconds on average), depending on the video duration. This corresponds to 135.6 to 636.7 seconds per configurator (258.4 seconds on average). Compared to manual usability analysis, these results indicate that MLLM-based analysis can be performed more efficiently.

\begin{figure}[h]
\vspace{-8pt}
\centering
\includegraphics[width=0.87\linewidth]{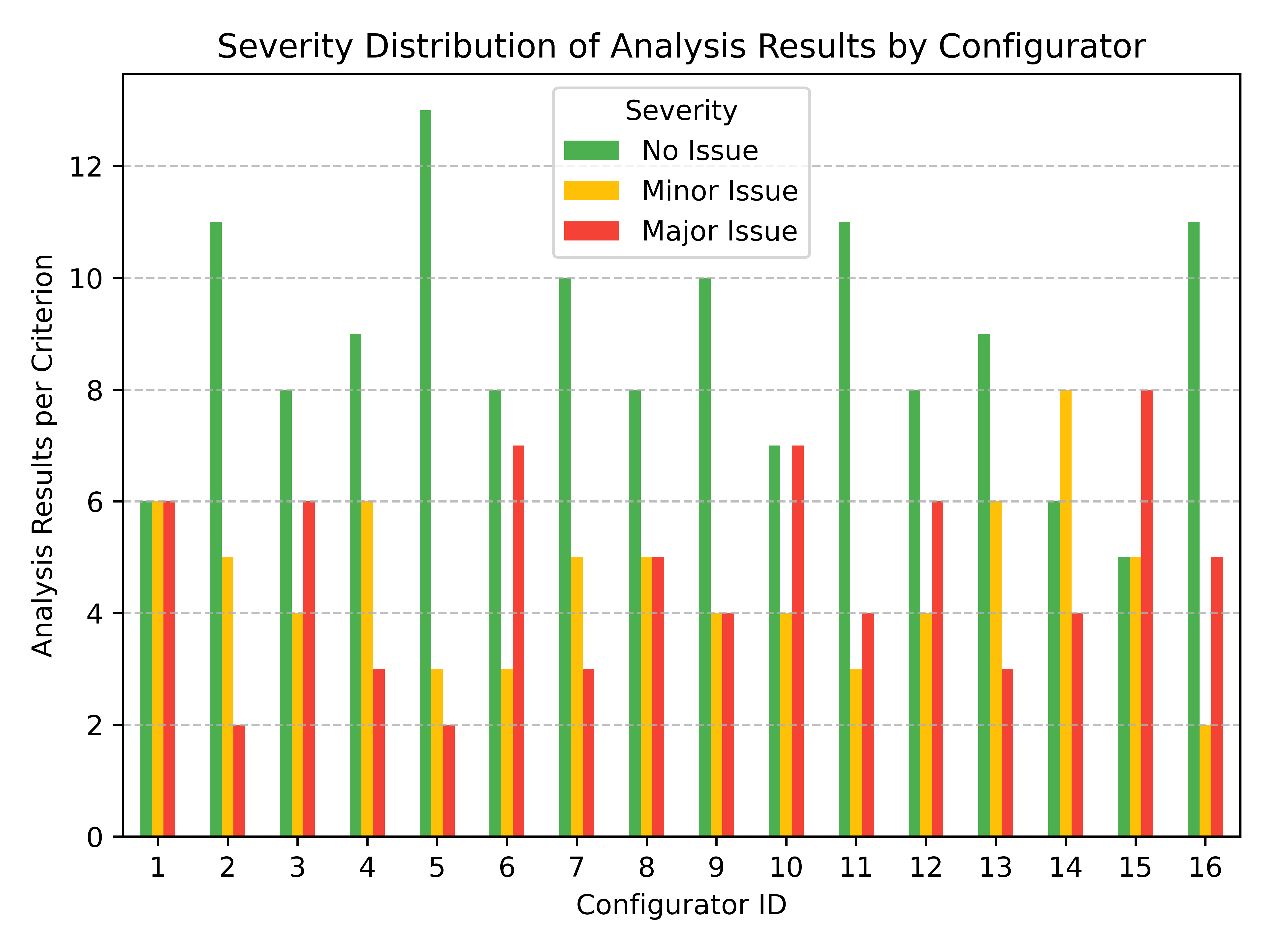}
\caption{Identified issues per configurator across all usability criteria.}
\label{fig:configurator_severity}
\vspace{-8pt}
\end{figure}

Across configurators, the number of identified issues (minor and major issues) ranged from 5 to 13 (6.75 on average). Figure~\ref{fig:configurator_severity} visualizes the distribution of issue severity across all analyzed usability criteria for each configurator. These results indicate that usability issues are common in configurator UIs. Additionally, the number of issues varies across configurators. This means that the MLLM considers the differences between configurators and provides individual results.

\begin{figure}[h]
\vspace{-8pt}
\centering
\includegraphics[width=0.87\linewidth]{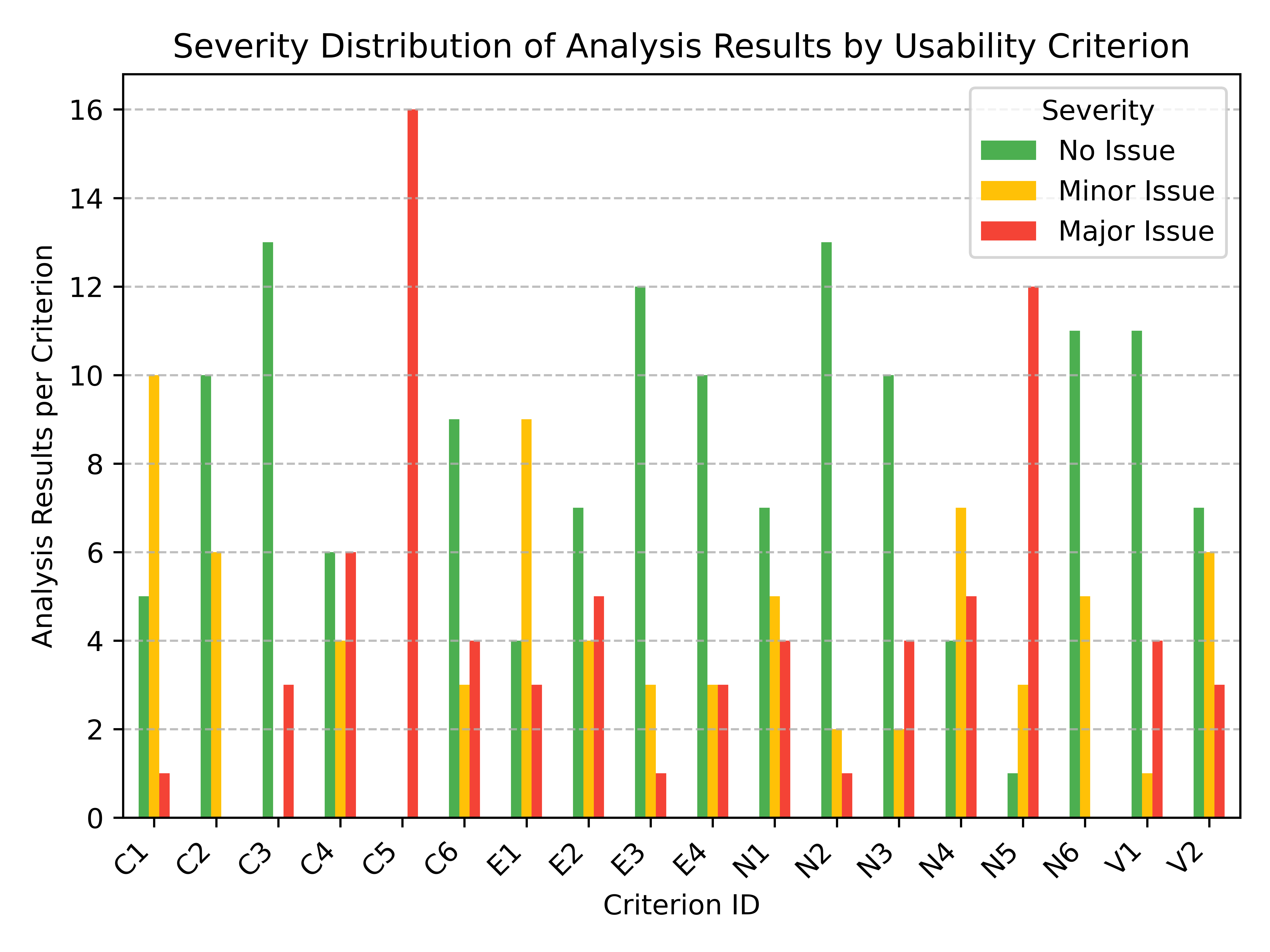}
\caption{Identified issues per usability criterion across all configurators.}
\label{fig:criterion_severity}
\end{figure}

Figure~\ref{fig:criterion_severity} summarizes how often issues were identified for each usability criterion. Certain criteria were fulfilled by most configurators, including \textit{availability of options (C3)}, \textit{transparency of errors (E3)}, \textit{manual step transitions (N2)}, \textit{starting points (N6)}, and \textit{product preview (V1)}. This indicates that these usability aspects are well established among configurator providers.

However, other criteria were frequently violated across configurators, especially \textit{variant comparison (C5)} and \textit{variant persistence (N5)}. These closely related criteria appear to be commonly overlooked. It is possible that product comparison is often not perceived as part of the core configuration task. Nevertheless, improving this aspect could enhance the user experience.

All analysis results in which the model reported \textit{minor} or \textit{major} issues were reviewed to evaluate their plausibility as described in Section~\ref{ssec:study_design}. To assess whether the plausibility assessment was consistent across the three reviewers per issue, we computed the \textit{inter-rater reliability} and report the results in Table~\ref{tab:irr_results}. The \textit{observed agreement ($P_o$)}~\cite{viera2005understanding} between reviewers was high and indicated substantial agreement. However, the \textit{Fleiss's Kappa} metric~\cite{fleiss1971kappa}, which accounts for chance agreement, indicates \textit{poor} agreement ($\kappa \approx 0$). The reason for this discrepancy is the \textit{Kappa Paradox}~\cite{konstantinidis2022kappaparadox}, in which extremely skewed ratings toward one option inflate the expected agreement by chance. The majority vote in our evaluation leads to 88.5\% of issue descriptions considered plausible and 98.0\% of improvement recommendations (see Table~\ref{tab:overall_results}). Therefore, minor disagreement is heavily affecting the \textit{Fleiss's Kappa} values. Consequently, we also report \textit{Gwet's AC1}~\cite{gwet2008ac1}, which is more robust to such skewed distributions and treats extremely dominant choices as intentional rather than by chance. These values indicate \textit{moderate} agreement ($0.4 <\kappa \leq 0.6$) for issue descriptions and \textit{almost perfect} agreement ($\kappa > 0.8$) for improvement recommendations.

\begin{table}[h]
    \centering
    \caption{Inter-Rater Reliability metrics for the review of MLLM-generated usability issue descriptions and improvement suggestions.} 
    \label{tab:irr_results}
    \vspace{4pt}
    \begin{tabular}{@{}lccc@{}}
    \toprule
    & \makecell{\textbf{Issue}\\\textbf{Description}} & &\makecell{\textbf{Improvement}\\\textbf{Recommendation}}\\ 
    \midrule
    \textit{Observed Agreement ($P_o$)} & 0.694 & & 0.842 \\
    \textit{Fleiss' Kappa} & 0.06 & & -0.007 \\ 
    \textit{Gwet’s AC1} & 0.546 & & 0.813 \\
    \bottomrule
    \end{tabular}
\end{table}

Overall, the majority of issue descriptions and improvement recommendations were considered plausible based on a majority vote as reported in Table~\ref{tab:overall_results}. The plausibility values did not show a notable difference between the assessment of \textit{major} and \textit{minor} issues. Full agreement among all three reviewers on the same judgment was observed for 54.1\% of the issue descriptions and 76.6\% of the improvement recommendations. 

\begin{table}[h]
    \centering
    \caption{Overall plausibility of MLLM-generated usability issue descriptions and improvement suggestions based on majority voting of reviewers.} 
    \label{tab:overall_results}
    \vspace{4pt}
    \begin{tabular}{@{}lccc@{}}
    \toprule
    & \makecell{\textbf{Issue}\\\textbf{Description}} & & \makecell{\textbf{Improvement}\\\textbf{Recommendation}}\\ 
    \midrule
    \textit{Minor Issues ($n=73$)} & 0.877 & & 0.973 \\
    \textit{Major Issues ($n=75$)} & 0.893 & & 0.987 \\
    \textit{All Issues ($n=148$)} & 0.885 & & 0.980 \\
    \bottomrule
    \end{tabular}
\end{table}

In the following, we discuss different usability analysis results in detail to further develop an understanding of their usefulness and current limitations. In our experiments, the MLLM consistently distinguished the severity of identified issues across all samples. If a criterion was only partially fulfilled, it was marked as \textit{minor issue}. In contrast, if the criteria were completely unfulfilled, they were marked as \textit{major issues}. Although this behavior is reasonable, assigning severity levels remains inherently subjective, as it depends on individual judgment rather than strictly objective criteria. Therefore, these labels should be used as a guide for grouping issues, not as a strict prioritization for issue resolution.

\begin{table}[h!]
    \centering
    \caption{Example analysis results of configurator UIs and usability criteria, including issue severity, description, and improvement suggestions.}
    \label{tab:analysis_results}
    \vspace{4pt}
    \begin{tabular}{@{}ccp{0.9cm}p{5.0cm}p{5.0cm}@{}}
    \toprule
    \rotatebox{90}{\textbf{Configurator}} & \rotatebox{90}{\textbf{Criterion}} & \rotatebox{90}{\textbf{\textbf{Severity}}} & \textbf{Description} & \textbf{Suggested Improvement} \\
    \midrule
    3 & V1 & \textit{Major issue} & \textit{The configurator lacks a continuous product preview that updates dynamically as the user makes selections regarding their hair characteristics (e.g., length, texture, oiliness, condition). The product image only appears towards the end of the process for naming, and the full personalized routine is only displayed on the final results page, significantly delaying visual feedback on the customization.} & \textit{Integrate a dynamic product preview that visually reflects the user's selections in real-time or near real-time throughout the questionnaire. For example, as the user selects hair goals, the product bottles could visually highlight the chosen benefits or ingredients, or show a representation of the personalized formula evolving. This would enhance engagement and help users understand how their choices impact the final product.} \\
    4 & N5 & \textit{Major issue} & \textit{The configurator lacks any visible functionality for users to save, name, or restore different configurations. While an 'Edit' button in the cart allows returning to the configurator, it doesn't support managing multiple saved designs or a version history, which is crucial for a product with extensive customization options.} & \textit{Implement a 'Save Design' feature within the configurator, allowing users to store their custom configurations. Provide a 'My Saved Designs' section where users can view, name, edit, or load previously saved variants. Consider offering options to share or duplicate designs.} \\
    8 & E1 & \textit{Minor issue} & \textit{While detailed information is available for accessories via 'i' icons, the initial 'Board' options (e.g., 'With extended cable space,' 'With space for cables') and the 'Add LED light with epoxy diffusion' checkbox lack in-context explanations or tooltips. Users must infer the differences from subtle visual changes or prior knowledge, which can lead to uncertainty.} & \textit{Add brief tooltips or small 'i' icons next to the 'Board' options and the 'Add LED light with epoxy diffusion' checkbox. These should provide concise explanations of what each option entails, its benefits, or key characteristics (e.g., 'Epoxy diffusion creates a soft, even light distribution').}\\
    \bottomrule
    \end{tabular}
    \vspace{32pt}
\end{table}

Table~\ref{tab:analysis_results} exemplifies some descriptions and suggested improvements of analysis results. Overall, these show that the generated issue descriptions were mostly nuanced and tailored to the specific configuration domain. They mention specific details of the configurators, like ``\textit{the ’Add LED light with epoxy diffusion’ checkbox lacks in-context explanations or tooltips}'' which allows users to easily locate the issue. Also, the descriptions reflect specific usability criteria, for example, ``\textit{it doesn’t support managing multiple saved designs or a version history,}'' which is clearly related to the \textit{variant persistence (N5)} criterion.

The suggested improvements directly address the described issues and provide explicit instructions on how to resolve them, for example, ``\textit{implement a ’Save Design’ feature within the configurator}'' or ``\textit{add brief tooltips or small ’i’ icons.}'' These recommendations specify concrete UI changes and also explain the underlying usability rationale, such as ``\textit{this would enhance engagement and help users understand how their choices impact the final product.}'' In this sense, the suggestions do not only mention surface-level observations, but also express design interventions that can be implemented by developers.

A review of the MLLM-generated analysis outputs that were considered implausible showed several common reasons that could guide future improvements of the analysis approach:

\begin{itemize}
    \item \textit{Unrecognized or misinterpreted controls}: Sometimes, the MLLM failed to correctly recognize or interpret interface elements shown in the screen recordings. For example, an icon-only control for saving configuration variants was not recognized as such, and subtle changes in the product visualization were overlooked. This led to incorrect or unsupported conclusions.
    \item \textit{Misinterpreted criteria}: In some cases, the MLLM did not correctly apply the intended meaning of a usability criterion. For instance, an issue was reported for \textit{error prevention (C6)} in the sense that incompatibility warnings were missing. However, the reason was that there were no incompatibilities in the example. This could be improved by extending the criteria descriptions with more detailed information on what should be observed.
    \item \textit{Overly strict interpretations}: In certain cases, the MLLM applied the criteria too literally and did not sufficiently include the configurator context. This was most evident for \textit{auto-completion (C4)} and \textit{progress indication (N4)}. For example, the MLLM identified missing auto-completion buttons as an issue, even though users could add the configuration to the cart at any time and any remaining properties were automatically filled with default values, which aligns with the intended meaning of auto-completion. Similarly, the MLLM identified a missing progress bar as an issue, though the configurators indicated progress through step indicators. Adding more detail or examples to the criteria definitions could help to reduce such rigid interpretations.
    \item \textit{Workflow mismatch}: The MLLM generally applied the criteria more accurately to configurators that customize a single product, while the accuracy was weaker for configurators that filter multiple products to identify the best match for a user’s needs. In these cases, especially issues related to the \textit{product preview (V1)} were incorrectly identified. Adapting the analysis instructions to better account for different configurator workflows could help mitigate this limitation.
\end{itemize}

These observations also help to explain the counterintuitive finding that some improvement recommendations were considered plausible, even though the issue description was implausible (see Table~\ref{tab:overall_results}). In these instances, the MLLM did not correctly recognize controls or interpreted the criteria too strictly, which means it did not capture a genuine issue in the description. Yet, the improvement recommendation was plausible and aligned with the criterion, such that it could be applied to further improve usability.

Overall, our evaluation suggests that MLLMs offer promising support for usability analysis of configurators, as they effectively interpret screen recordings, apply domain-specific criteria, and generate detailed, context-aware descriptions and improvement recommendations. While further evaluations are still required, as not all results accurately describe genuine issues, the presented approach can substantially reduce evaluation effort.

\section{Discussion} \label{sec:discussion}
Our results demonstrate that MLLMs can meaningfully enhance the usability analysis of configurators by identifying diverse issues with relatively low effort. The model generated domain-aware explanations, provided differentiated assessments across systems, and offered explicit suggestions for improvement. This indicates that configurator-specific usability criteria can be analyzed using multimodal input context, which creates promising opportunities for integrating automated analysis into real-world development workflows.

From a practical perspective, the findings highlight several implications for improving configurator UIs. Firstly, attention can be directed to frequently overlooked aspects, such as \textit{variant comparison} and \textit{variant persistence}. The automated detection can help providers identify these recurring gaps early and encourage them to provide additional support for users. Secondly, the detailed recommendations generated by the model suggest actionable refinements, such as adding clarifying tooltips, enhancing visual feedback, or providing clearer progress indicators. Such guidance can accelerate iteration cycles, especially in teams without dedicated usability experts.

Additional development is needed to further reduce implausible outputs. This includes refining prompt instructions, for example, by providing few-shot examples~\cite{reynolds2021prompt}, specifying criteria more precisely with explicitly observable measures, and adapting to different configurator workflows more effectively. Structured improvements in these areas, combined with systematic evaluation, could reduce the number of false positives and further improve the quality of results.

Automated usability analysis of configurators using MLLMs has the potential to create a substantial practical impact. These tools could continuously monitor usability throughout design and development, providing rapid feedback in each iteration to identify overlooked issues early. Keeping human validation in the loop ensures that automatically identified issues are interpreted correctly and that configurator quality is not compromised by false positives or misaligned recommendations. 
A current boundary is the need to manually generate screen recordings. While this approach still reduces the effort of the actual usability analysis, the use of intelligent agents to simulate user behavior and generate recordings could further automate the evaluation~\cite{yoon2024intent}. However, these approaches are not yet accurate enough for controlled study settings.


While our current approach focuses on textual improvement suggestions, these improvements could be accompanied by concrete UI code changes if a configurator’s code repository is accessible~\cite{jiang2026survey}. This way, the MLLM acts as an assistant that translates usability findings into implementation-level guidance to support developers in a semi-automated way.

Overall, the results suggest that MLLM-based usability analysis can evolve from feasibility to practical utility, enabling more accessible, scalable, and systematic improvements to configurator UIs.

\section{Threats to Validity} \label{sec:threats}
Our analysis relies on 18 configurator-specific usability criteria derived from related studies. However, there may be additional relevant criteria that were missed in our literature review, for which the analysis does not work as expected. We attempted to mitigate this issue by employing a broad, diverse range of criteria aggregated from related literature.

The screen recordings used as input represent simulated rather than real user interactions. Real interactions could raise other issues during analysis. We tried to mitigate this aspect by demonstrating a broad range of potential user interactions in the recordings, which should provide sufficient context for the analysis. Nevertheless, future studies should validate the results with interaction recordings of a more representative number of users.

An assessment of completeness, i.e., whether all existing issues in the configurators were detected, was beyond the scope of this work, which focused on the correctness of the identified issues. Nevertheless, completeness is a complementary aspect for understanding the extent to which MLLMs identify usability issues compared to usability experts. We therefore plan to address this aspect in future work by comparing MLLM outputs with additional evaluations conducted by usability experts.


Finally, the study evaluated 16 configurators of different industries. Although this diverse sample provides a broad range of examples, it remains limited in size and may not capture the full variability of configurator designs, interaction paradigms, or domain conventions. A broader dataset would strengthen external validity and support more robust generalization of the findings.

\section{Conclusions} \label{sec:conclusions}
This work investigated how \textit{multimodal large language models (MLLM)} can support the usability analysis of configurators. Building on prior work, we derived 18 configurator-specific usability criteria and applied them to 16 real-world configurators using an MLLM and screen interaction recordings as contextual input. The model identified multiple issues per configurator. A review involving multiple software engineers found that most reported issues were plausible and accompanied by useful, domain-specific improvement suggestions. 

Practically, our results indicate that MLLMs offer a low-effort approach to analyzing the usability of configurators, although a follow-up expert-based validation remains necessary. With further refinement, such analyses could be integrated as continuous, automated checks in configurator engineering workflows. Our future work includes refining the criteria and prompts to further reduce the detection of implausible usability and further improve result quality. We also plan to expand the scope of our evaluations to better account for aspects of completeness and integration in software development processes.



%
%
%
\bibliographystyle{splncs04}
\bibliography{bibliography}






\end{document}